\documentclass[preprint2]{emulateapj}
\shortauthors{Xing et al.}

\newcommand{\fermi}{\textit{Fermi}}

\begin{document}

\title{The likely \textit{Fermi} Detection of the Supernova Remnant RCW 103}

\author{Yi Xing\altaffilmark{1}, Zhongxiang Wang\altaffilmark{1},
Xiao Zhang\altaffilmark{2}, \& Yang Chen\altaffilmark{2,3}}

\altaffiltext{1}{\footnotesize 
Key Laboratory for Research in Galaxies and Cosmology,
Shanghai Astronomical Observatory, Chinese Academy of Sciences,
80 Nandan Road, Shanghai 200030, China}

\altaffiltext{2}{\footnotesize Department Astronomy, Nanjing University,
22 Hankou Road, Nanjing 210093, China}

\altaffiltext{3}{\footnotesize 
Key Laboratory of Modern Astronomy and Astrophysics,
Nanjing University, Ministry of Education, Nanjing 210093, China}

\begin{abstract}
We report on the results from our $\gamma$-ray analysis of
the supernova remnant (SNR) RCW~103 region. The data were taken with
the Large Area Telescope on board the \textit{Fermi Gamma-ray Space Telescope}.
An extended source is found at a position consistent with that 
of RCW~103, and its emission
was only detected above 1~GeV (10$\sigma$ significance), 
having a power-law spectrum with a photon index 
of 2.0$\pm$0.1. We obtain its 1--300 GeV spectrum, and the total flux
gives a luminosity of 8.3$\times 10^{33}$ erg s$^{-1}$ at a source distance
of 3.3 kpc. 
Given the positional coincidence and property similarities
of this source with other SNRs, we identify it as the 
likely \textit{Fermi}
$\gamma$-ray counterpart to RCW~103. Including radio measurements of RCW~103,
the spectral energy distribution (SED) is modeled by considering
emission mechanisms based on both hadronic and leptonic scenarios. 
We find that models in the two scenarios can reproduce the observed SED, 
while in the hadronic scenario the existence of SNR--molecular-cloud 
interaction is suggested as a high density of the target protons is 
required.
\end{abstract}


\section{Introduction}

The properties of the supernova remnant (SNR) RCW 103 (G332.4-0.4) 
has been studied at multiple energies, and the SNR is well known as 
it contains an enigmatic central compact object (CCO; 1E 161348$-$5055, 
hereafter 1E~1613). Having a size of $\sim$10\arcmin\ in 
diameter \citep{cas+80,tg80}, it was determined from optical imaging to 
have a shell expansion rate of 
1100 km~s$^{-1}$ \citep{cdb97} for a source distance of 
3.3 kpc \citep{cas+75}. This expansion rate implies an age of 
approximately 2000 yrs. The mid- and near-infrared property 
characteristic of molecular shock, 
the nearby H$_2$ emission, and the HCO$^{+}$ morphological feature 
suggests that it is interacting with a molecular cloud 
(see \citealt{jia+10} and references therein).
Although it appears similar to typical CCOs by being radio-quiet and 
not having non-thermal point-source and extended emission 
(for detailed properties of CCOs, 
see \citealt{pst04}; \citealt{del08}; and more recently \citealt*{gha13}), 
the X-ray point source located in the center of RCW~103 \citep{tg80}
shows strong X-ray variability \citep{gpv99}
and has an X-ray periodicity of 6.67 hr \citep{del+06, esp+11},
making itself unique among known young neutron stars. 
The properties of this young neutron star is 
poorly understood, and different possibilities have been proposed 
\citep{lix07,piz+08,bg09,ikh+13}.

SNRs are known to have high-energy non-thermal emission, arising from 
the shocks of SN explosions.  With the great capabilities of the 
\textit{Fermi Gamma-ray Space Telescope}, many SNRs have been detected at 
its GeV $\gamma$-ray energies. Several of them are known 
to contain CCOs, and they are Cassiopeia A (Cas A; \citealt{aaa+10}), 
Vela Jr. (G266.2$-$1.2; \citealt{tan+11}),
Puppis A (Pup~A; \citealt{hew+12}), and 
PKS 1209$-$51/52 (G296.5+10.0; \citealt{ara13}).
Similar to other SNRs at the GeV energy range, these young
SNRs that harbor a CCO generally 
have extended power-law emission with photon indices of $\sim 2$.
With the current \fermi\ measurements, both a leptonic or a
hadronic scenario can describe the observed broad-band spectra, while
for individual sources one of the scenarios may be slightly more favored
(see, e.g., \citealt{ara13} and references therein).
No indication of GeV emission from the CCOs has been found; note that
for the CCOs in Pup~A and PKS~1209$-$51/52, their spin periods
are known from X-ray timing (\citealt{gha13} and references therein).

In this paper we present our analyses of the \fermi\ data of the RCW~103
region, and report the likely detection of its GeV emission.
In Section 2 the \fermi\  observations are described, and in
Section 3 different data analyses and results are given. 
We discuss our results in Section~4.

\section{Observations}
\label{sec:obs}

The Large Area Telescope (LAT) is a $\gamma$-ray imaging instrument on board
the \textit{Fermi Gamma-ray Space Telescope}, which continuously scans 
the whole sky every three hours in energy range from 
20 MeV to 300 GeV \citep{atw2009}. In our analyses we selected LAT events 
inside a 20$\arcdeg\times 20\arcdeg$ region centered at the position of 
the SNR RCW 103 from the \textit{Fermi} Pass 7 database. 
The time period of the data is from 
2008-08-04 15:43:36 (UTC) to 2013-09-09 00:40:00 (UTC).
We rejected events below 200 MeV 
because of the relative large uncertainties of the instrument response 
function of the LAT in the low energy range. In addition we only 
included events 
with event zenith angles less than 100 degrees to prevent the Earth's limb 
contamination, and during good time intervals when the quality of the data 
was not affected by the spacecraft events. These selections are recommended by 
the LAT team.
\begin{center}
\includegraphics[scale=0.35]{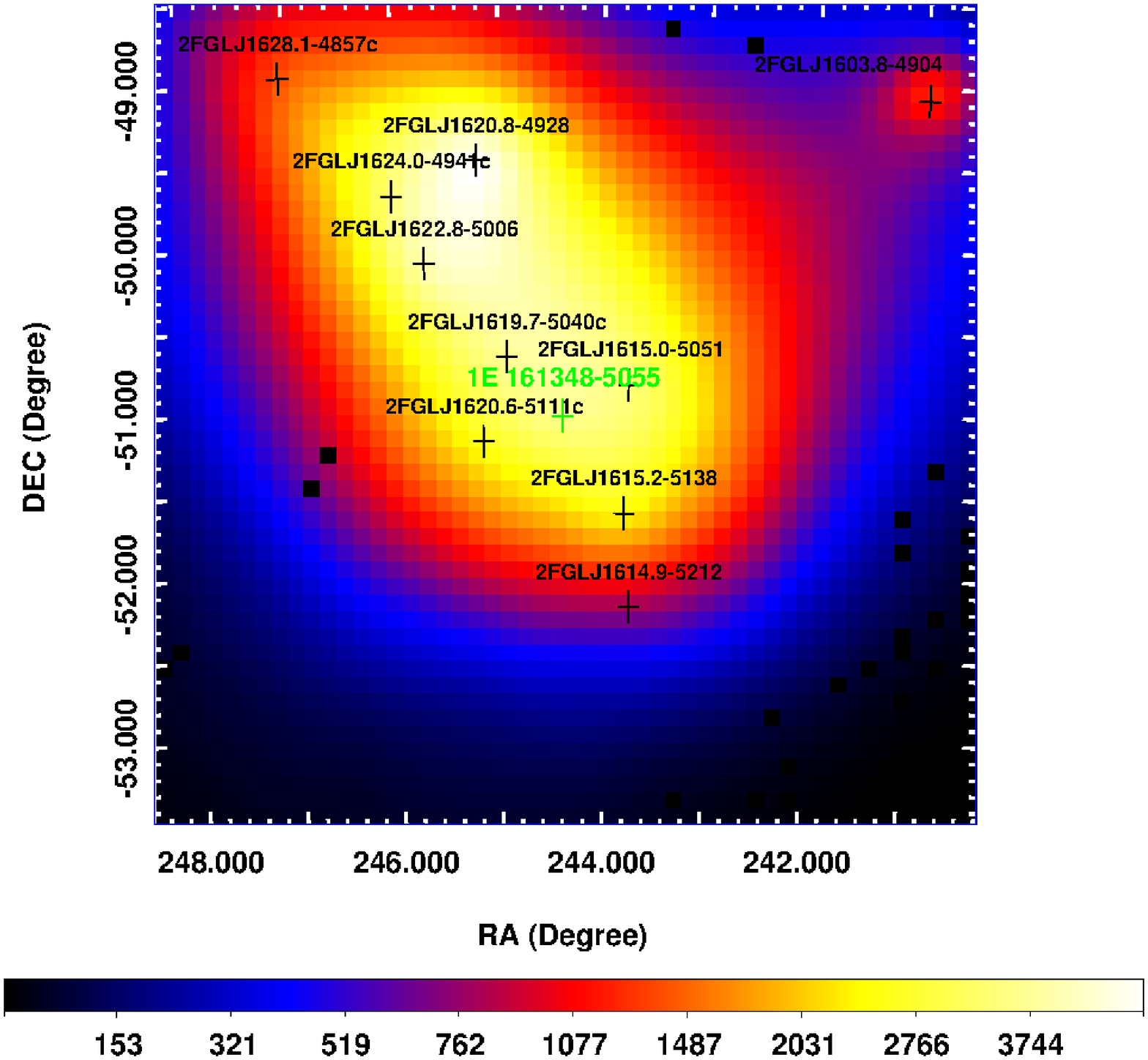}
\includegraphics[scale=0.35]{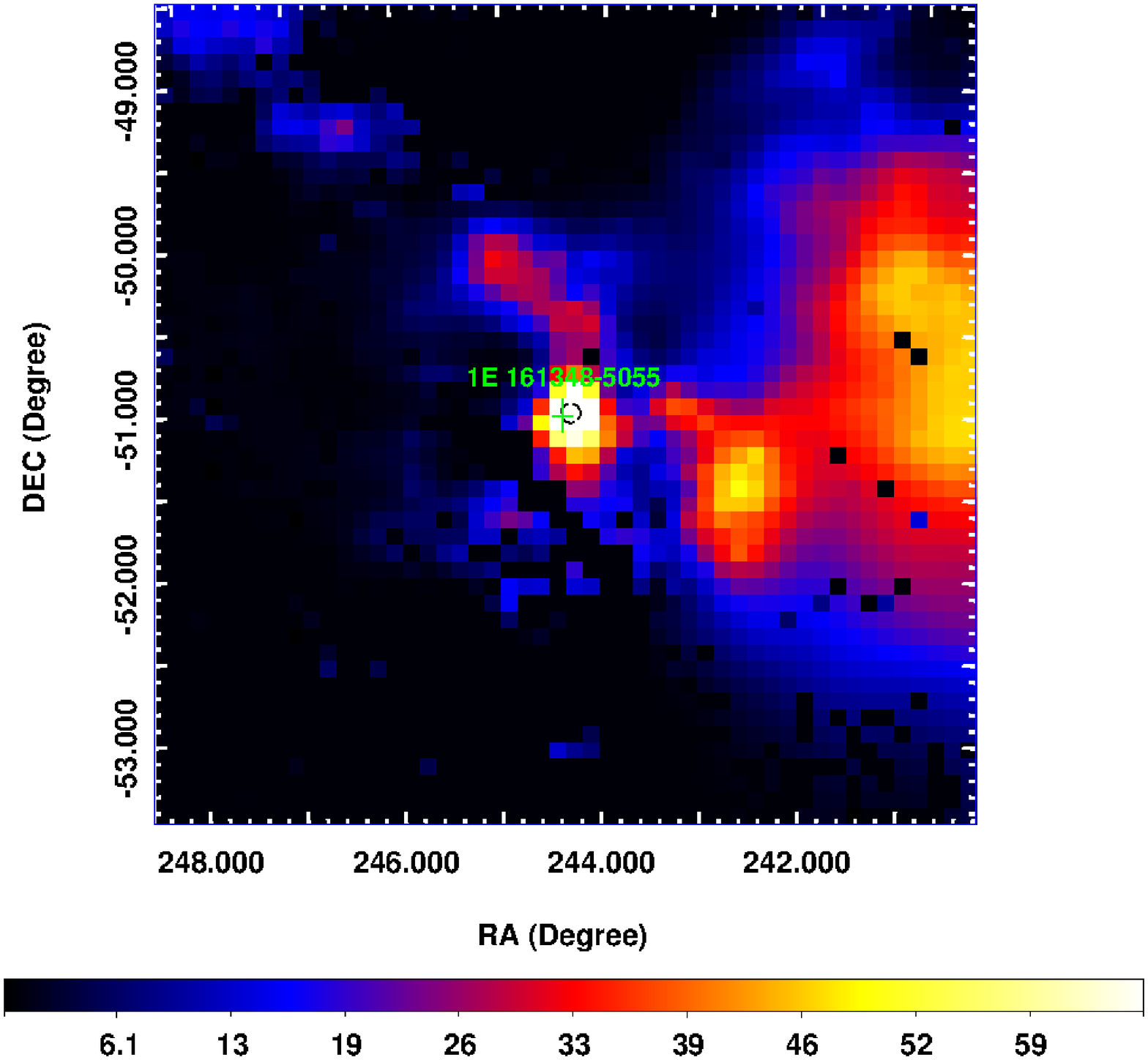}
\figcaption{200 MeV$-$300 GeV TS maps of the $\mathrm{5^{o}\times5^{o}}$ 
region centered at RCW 103. The image scales of the maps are 
0.1\arcdeg\ pixel$^{-1}$. {\it Upper panel}: sources in 
the source model outside of the region were considered, with sources
in the \fermi\ 2-year catalog within the region marked.
{\it Bottom panel}: all sources in the source model were considered. 
The dashed circle indicates the 2$\sigma$ error circle of the best-fit 
position for the residual emission found at the position of RCW~103.
\label{fig:lmap}}
\end{center}

\section{Analysis and Results}

\subsection{Source Identification}
\label{subsec:si}

We first included all sources within 15 degrees centered at the position of 
RCW 103 (CCO's position: R.A.=16$^{\rm h}$17$^{\rm m}36\fs3$, 
Decl.=$-$51\arcdeg02\arcmin24\farcs5, equinox J2000.0; \citealt{pst04})  
in the \textit{Fermi} 2-year catalog \citep{nol+12}
to make the source model. 
The spectral function forms of these sources are provided in the catalog. 
We let the spectral normalization parameters of the sources 
within 4 degrees from RCW 103 free, and fixed all the other parameters 
of the sources to their catalog values. We also included the spectrum model 
gal\_2yearp7v6\_v0.fits and the spectrum file iso\_p7v6source.txt in 
the source model to consider the galactic and the extragalactic diffuse 
emission, respectively. The parameters `Value' of the Galactic diffuse 
emission model and `Normalization' of the extragalactic diffuse 
emission model were let free.
We performed standard binned likelihood analysis to the LAT data with the 
LAT science tools software package {\tt v9r31p1}, and extracted the 
Test Statistic (TS) map of a $5\arcdeg\times 5\arcdeg$ region centered 
at the position of RCW 103. 
A source map considering sources in the source 
model outside of the region was made, which is shown in the upper panel of
Figure~\ref{fig:lmap}.  
It can be seen from the TS map that RCW~103 
is located in a very complex region. 

After considering and removing all the sources in the source model in 
this region, we then made a residual map, which is shown in the bottom 
panel of Figure~\ref{fig:lmap}.  As can be seen, excess $\gamma$-ray emission 
remained near the center, and TS$\simeq$60, indicating $\sim$8$\sigma$ detection
significance.  We ran \textit{gtfindsrc} in 
the LAT software package to find the best-fit position of 
the excess $\gamma$-ray emission
and obtained a position of R.A.=244\fdg319, Decl.=$-$51\fdg0261,
(equinox J2000.0), with 1$\sigma$ nominal uncertainty 
of 0\fdg03. In addition, detailed analysis indicated that the excess
emission only appeared above 1~GeV, as the TS value at the region
was nearly zero when only the energy range of 0.2--1~GeV was used. 
A TS map was thus made with $\geq$1~GeV photons from the RCW~103 region
and a region of $1\arcdeg\times 1\arcdeg$ centered at RCW~103
is shown Figure~\ref{fig:smap}.  The detection 
significance now is improved to $\simeq$10$\sigma$ (TS$>$100).
The CCO 1E~1613 centered at RCW 103 is located 
slightly outside of the 1$\sigma$ error circle 
with an angular separation of 0\fdg05, but within 
the 2$\sigma$ error circle. 

There are two nearby sources that could be associated with the excess
emission, which are PSR J1617$-$5055 and HESS J1616$-$508 \citep{hessgal06}.
\citet{lan+07} analyzed archival X-ray data
and suggested that the HESS source is the pulsar wind nebula (PWN) 
powered by J1617$-$5055. In Figure~\ref{fig:smap}, the pulsar's location
and the source size (16\arcmin\ diameter) of HESS J616$-$508 are marked. 
The pulsar is $\approx$3.7$\sigma$ away from our \fermi\ source,
and in \S~\ref{sec:dis} we argue that the \fermi\ source is not likely
the associated PWN on the basis of spectral property comparison and 
source positions. 

Both PWNe and SNRs are the main sources detected by the HESS
survey of the Galactic plane (see, e.g., \citealt{car+13}) at its TeV
energy range. \citet{fhj1837} searched through sources in 
the \fermi\ 2-year catalog and found that 2FGL J1615.0$-$5051 (see
the top panel of Figure~\ref{fig:lmap}) is extended 
and spatially coincident with HESS J1616$-$508, suggesting that
they are very likely associated (see also \citealt{ace+13}). 
In our analysis above, 2FGL J1615.0$-$5051 was treated as a point source,
which might not be appropriate if it is truely extended
(note that because of source crowdedness, contamination between the sources
can not be avoided).
We tested to include 2FGL J1615.0$-$5051 as an extended source 
(0.32 deg size; \citealt{fhj1837}) in the source model, and found that excess
emission was still detected at the same position but with TS$\simeq$40.
We further checked the fit improvement by 
calculating
the significance values (estimated from $\sqrt{2\log(L_2/L_1)}$, 
where $L$ is the likelihood
value; e.g., \citealt{fhj1837}) for different setups of 2FGL J1615.0$-$5051 and 
the new \fermi\ source. In the calculation, the model for $L_1$ only
had the extended source given by \citet{fhj1837}, and the models for
$L_2$ had 2FGL J1615.0$-$5051 plus the new \fermi\ source, both being either
a point source or an extended source (for the latter case 
for the new \fermi\ source, a uniform disk with a radius of 0\fdg3 was
used; see below \S~\ref{subsec:sda}). We found that the lowest
significance value was 5.5 when 2FGL J1615.0$-$5051 and the new \fermi\ source
were the extended source and a point source, respectively, and 
the highest value was 10.2 when the first and the latter were a point
source and an extended source, respectively. The analyses indicate that not only
the new \fermi\ source was clearly detected but also a point source
2FGL J1615.0$-$5051 is more favored.
\begin{center}
\includegraphics[scale=0.35]{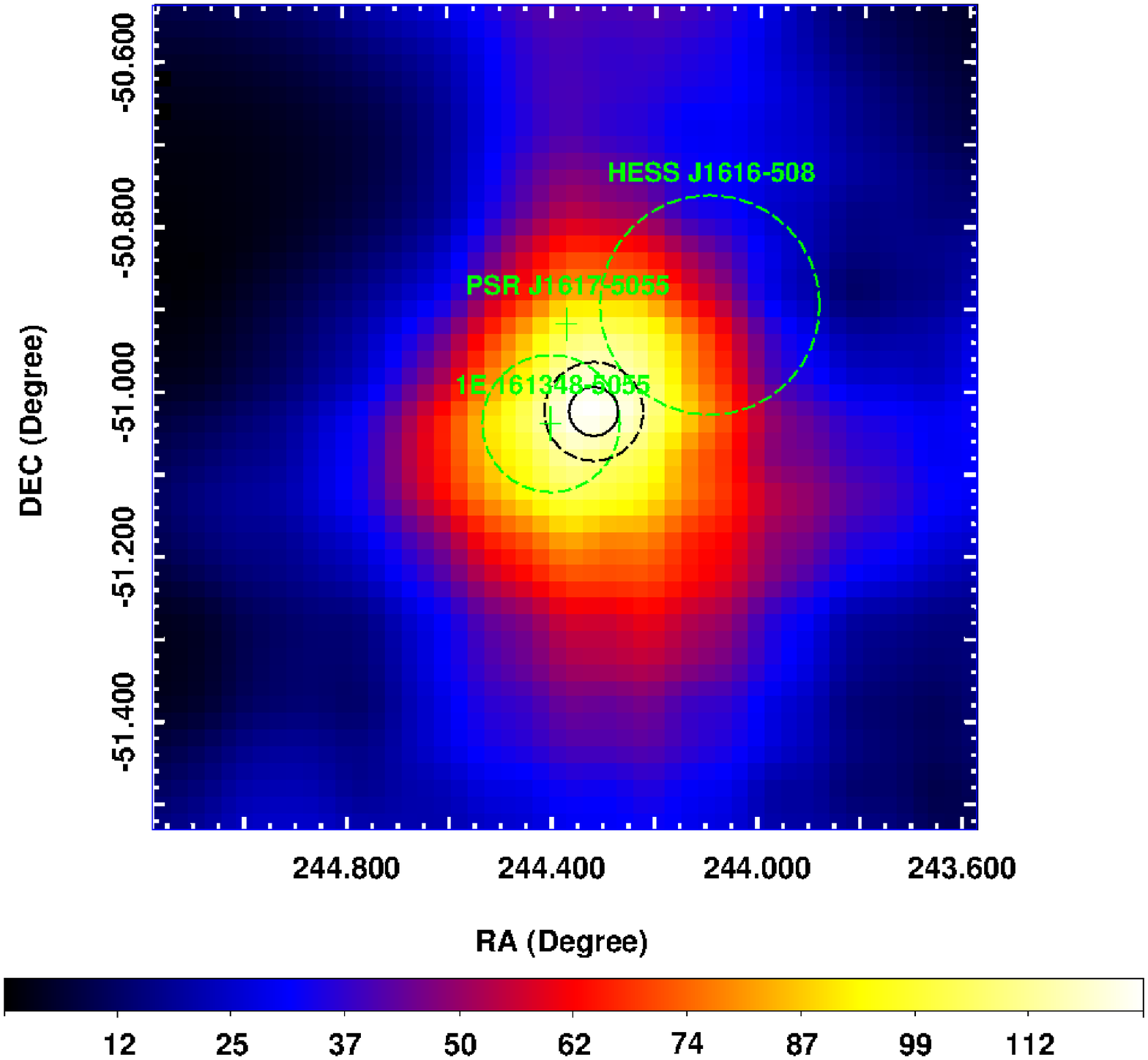}
\figcaption{Residual 1$-$300 GeV TS maps of $1\arcdeg\times 1\arcdeg$ 
region centered at RCW 103. The image scale of the maps 
is 0\fdg02 pixel$^{-1}$.  
The \fermi\ source's 1$\sigma$ and 2$\sigma$ positional error circles 
are marked by black solid and dashed circles, respectively,
the RCW~103 region centered at the CCO is marked
by a green dashed circle, and the positions of
PSR J1617$-$5055 and HESS J1616$-$508 are also marked, with the latter
indicated by the larger green dashed circle. 
\label{fig:smap}}
\end{center}

\subsection{Spatial Distribution Analysis}
\label{subsec:sda}

We analyzed the spatial distribution of the new \fermi\  $\gamma$-ray source 
at RCW~103 to determine whether the excess emission is point-like or 
extended. 
We used both a point source with a power-law spectrum
at the best-fit position and uniform disk models with power-law spectra
to analyze the emission in the 30$-$300 GeV range. The searched radius range
for the uniform disks was 0\fdg1--0\fdg5 (see Table~\ref{tab1}), 
and the high energy range
was used for the optimal spatial resolution. Additionally in the analysis, 
only front converting events for the instrument response function 
P7SOURCE\_V6::FRONT were included, which allows to reduce the point-spread
function (PSF) of the LAT to $<$0\fdg15 (68\% containment).
We fixed the power-law indices at 2 for the models 
(obtained from likelihood analysis in $>$1~GeV energy range; 
see below \S~\ref{subsec:sa}) to reduce the uncertainties. 
For the point source, we let the spectral normalization parameters of the 
sources within 4 degrees from RCW 103 free, and fixed all the other 
parameters of the sources in the source model at the \fermi\ 2-year 
catalog values (2FGL J1615.0$-$5051 was included as a point source
on the basis of our analyses above in \S~\ref{subsec:si}).
For the disk models, we fixed all spectral parameters of the sources 
in the source model at the values obtained above, but let the spectral 
normalization parameters of the disk models free.
We obtained a TS of 24 for the point source model and 
a maximum TS value of 37.5 at the radius of 
0\fdg3 for the disk models, although we note that the TS values 
for the radius range of 0\fdg16--0\fdg3 do not indicate any 
significant differences (Table~\ref{tab1}). 
Comparing the TS values, our analysis implies 
$>$3$\sigma$ detection of the source extension
(the significance was calculated from 
$\sqrt{{\rm TS}_{disk}-{\rm TS}_{point}}$; see, e.g., \citealt{fhj1837}). 
The obtained photon fluxes for these models
are given in Table~\ref{tab1}.


\subsection{Spectral Analysis}
\label{subsec:sa}

From our likelihood analysis, the excess $\gamma$-ray emission was
found to be detected only above $\sim$1 GeV. Different source models with 
a power-law spectrum of $dN/dE=N_{0}E^{-\Gamma}$, which
included a point source or extended sources at the best-fit position, 
were added to the source model, and the emission was found to have 
$\Gamma$ of 1.9--2.0 with an uncertainty of 0.1.
Given the above results from \S~\ref{subsec:si} and \S~\ref{subsec:sda},
we report our $\gamma$-ray 
spectrum result by considering the excess $\gamma$-ray emission as an 
extended source with a size radius of 0\fdg3 at the best-fit position. 
The $\gamma$-ray spectrum was obtained by
performing maximum likelihood analysis to the LAT data 
in 5 evenly divided energy bands in logarithm from 1--300 GeV.
Similar to those of the other SNRs, the obtained spectrum has a 
relative flat energy 
distribution with a photon index of $\Gamma=2.0\pm0.1$. The energy
and flux values at the 5 bands are given in Table~\ref{tab2}.
The total 1--300 GeV
luminosity was 8.3$\times 10^{33} (D/3.3\ {\rm kpc})^2$ erg~s$^{-1}$,
where source distance $D=$3.3 kpc was used for RCW~103 \citep{cas+75}.

\subsection{Timing Analysis}

We performed timing analysis to the \textit{Fermi}/LAT data of 
the RCW~103 CCO region to search for any modulations. 
The LAT data within 0\fdg2 from the position of the
CCO 1E~1613 were folded at its 6.67 hr periodicity \citep{del+06,esp+11}, 
and two energy ranges, 0.2--300 GeV and 1--300 GeV, were respectively used.
No modulations at the period were 
detected. The values from the $H$ test \citep{dej94} obtained from 
the folded light curves were 
0.2 and 0.1 in the ranges of 0.2--300 GeV and 1--300 GeV, respectively, 
which are significantly small. 
The value of $H=42$ is used by the LAT team to 
confirm $\gamma$-ray pulsations \citep{abdo2010}. 

In addition we also constructed 1000~s binned light curves in
the above two energy ranges, which were obtained using 
\textit{Fermi}/LAT aperture photometry analysis. An aperture radius 
of 0\fdg2 was used. The power spectra in the two energy 
ranges were extracted. The exposures 
used to determine the flux in each time bins were calculated assuming 
a power-law spectrum with $\Gamma=2$. No modulations in the two energy 
ranges were detected. 

We tested to increase the radius used for epoch folding and aperture photometry 
analysis to 0\fdg4, but no modulation at the known period
or other periods were detected.
The obtained $H$ values for the folded light curves
were similarly small (0.5 and 1.9) as those given in the above 
in the two energy ranges.
\begin{center}
\includegraphics[scale=1.1]{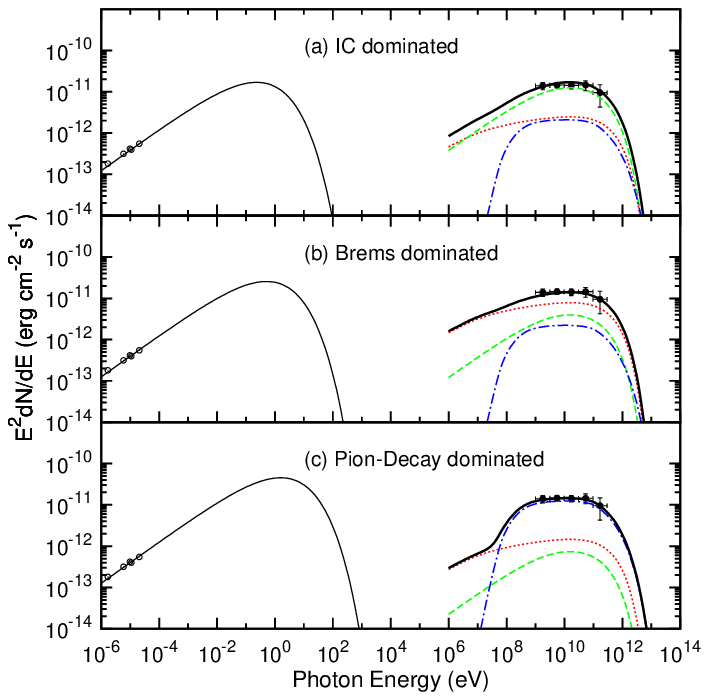}
\figcaption{\fermi\ $\gamma$-ray SED of RCW 103. Radio measurements
of the SNR are also included. Three emission components due to
IC scattering (green dashed curve), 
bremsstrahlung (red solid curve), and 
$\pi^0$ decay (blue dash-dotted curve) are combined to fit the SED.
\label{fig:sed}}
\end{center}

\section{Discussion}
\label{sec:dis}

Analyzing the \fermi/LAT data of the RCW~103 region,
we found an extended $\gamma$-ray source with $\sim$10$\sigma$ significance
at a position consistent with
that of the young SNR. It should be cautiously noted that
as shown in both Figures~\ref{fig:lmap} and \ref{fig:smap}, 
the source is located in a complex region. A few 
known \fermi\ $\gamma$-ray sources, the young pulsar J1617$-$5055,
and HESS J1616$-$508 are nearby. The intensity-peak position of
this \fermi\ source also appears to have a $\sim$0\fdg05 offset from 
the center of RCW~103 (however the intensity peak position 
roughly coincides with the north-west edge of the X-ray shell which 
overlaps one bright H$_2$ region; see \citealt{omd90}).
We determined that the \fermi\ source probably had a radius of as small as
$\sim$0\fdg16, which is approximately double the size of the SNR 
seen at X-ray and radio energies. However the property similarity 
of the source and the other SNRs strongly supports the detection 
of \fermi\ $\gamma$-ray emission from RCW~103. 
For example for those young SNRs harboring a CCO,
they all have 
prominent emission at energies above 1~GeV, and the spectra are a 
power law with photon indices in a range of 1.85--2.1, which makes their 
spectral energy distributions (SEDs) flat in the energy range. 
At similar distances, they have luminosities of 
10$^{33}$--10$^{34}$ erg s$^{-1}$. The detection of H$_2$ lines in
the region right outside of the remnant shell of RCW~103
suggests that the remnant is interacting, probably fractionally,
with a molecular cloud (\citealt{omd90}; see also \citealt{jia+10}). 
The `normal' $\gamma$-ray luminosity 
value we derived for RCW~103 is consistent with the picture, as the SNRs
that are known to be interacting with molecular clouds have luminosities
at least one order of magnitude higher because of 
the high target masses of molecular clouds (e.g., \citealt{abd+09,cs10}).

The pulsar J1617$-$5055 (having a spin-down age of 8.1 kyr; 
\citealt{tor+98,kas+98}) is located 3.7$\sigma$ away from the position
of the detected \fermi\ source. Since electrons responsible for 
$\gamma$-ray emission of PWNe via inverse Compton (IC) scattering
are thought to be `old' (i.e., they are less energetic and have longer
lifetimes than those X-ray emitting electrons detected around pulsars 
due to synchrotron radiation; e.g., \citealt{mat+09,dej+09}), 
a significant offset between a fast-moving pulsar and its GeV/TeV PWN might
appear (e.g., \citealt{krp13}).
However, the currently confirmed \fermi\ PWNe all have harder power-law spectra 
with photon indices in a range of 1.4--1.6 
(e.g., the Crab pulsar, \citealt{fcrab10}; 
PSR B1509$-$58, \citealt{fb1509}; 
PSR J1838$-$0655, \citealt{fhj1837};
PSR J1856+0245, \citealt{fhj1857};
PSR B1823$-$13, \citealt{fb1823}), making their SEDs clearly rising in 
the \fermi\ energy range.
The difference in the spectral properties of these PWNe and the SNRs is 
distinguishable. Moreover, even if the $\gamma$-ray source is a PWN powered
by J1617$-$5055, it would imply that the pulsar not only moved away
from a sky region coinciding with RCW 103, raising the issue again about
whether or not they are associated \citep{kas+98}, but also has to have an
extremely large transverse velocity ($\sim 4200$ km s$^{-1}$;
see \citealt{kas+98} for detailed discussion).  
The current studies of RCW~103 and the pulsar do not support either
of them.

We searched in the SIMBAD Astronomical
Database within the 2$\sigma$ error circle of the best-fit position of
the $\gamma$-ray source, but
only a few normal stars besides RCW~103 and its CCO are known in the region.  
Given all these, we conclude that
\fermi\ $\gamma$-ray emission from the SNR RCW~103 was likely detected, 
although contamination from nearby $\gamma$-ray sources due to the low
spatial resolution of the LAT is possible (Figure~\ref{fig:lmap}; 
a TS of $\sim$2000
at the position of RCW~103 when the nearby sources
are kept versus a TS of $\sim$60 in the residual map).

With the conclusion, we studied the SED of RCW~103 by considering both
the hadronic and leptonic scenarios. 
In the scenarios, a power-law
spectrum with a cut-off energy $E_{i,cut}$ for particles of electrons
and protons is assumed:
\begin{equation}
dN_i/dE_i=A_i E_i^{-\alpha_i} exp(-E_i/E_{i,cut}),
\end{equation}
where $i=e,p$, $E_i$ is the particle kinetic energy, $\alpha_i$
is the spectral index, and $A_i$ is the normalization factor.
A ratio of $K_{ep}=A_e/A_p$ compares the number of the electrons to that of
the protons at a given energy. We included the radio flux measurements of
the SNR \citep{bea66,gs70,sg70,cas+80,dic+96} as additional constraints, 
which can be described by a power law with a spectral index of $-0.56$ 
(Figure~\ref{fig:sed}; \citealt{dic+96}). In the hadronic 
scenario, $\gamma$-rays
are emitted due to the decay of $\pi^0$ mesons produced in collisions of
the protons with ambient gas, and in the leptonic scenario, IC scattering 
or bremsstrahlung emission by/from high-energy electrons
contributes dominantly to the observed $\gamma$-rays. 
We refer to \citet{zha+13} and references therein for calculation details.

We found that both scenarios can describe the SED. Our model
spectra that are dominated by IC scattering, bremsstrahlung, or
$\pi^0$ decay components are shown 
in the upper, middle, and bottom panel
of Figure~\ref{fig:sed}, respectively. In the calculations, an energy 
density of 0.5 eV cm$^{-3}$ for the interstellar radiation field at 
the location \citep{pms06} was used,  $\alpha_e=2.0$
was needed to fit the radio data points, and $\alpha_p$=2.0,
$E_{e,cut}=1$ TeV, and $E_{p,cut}=3$ TeV were found to be able to provide
a good fit to the $\gamma$-ray part. Our model fluxes at the X-ray 
energy range of 0.2--10 keV is generally below 
10$^{-14}$~erg cm$^{-2}$ s$^{-1}$,
which may explain the non-detection of a power-law component in the
X-ray spectrum of the SNR \citep{nug+84,gph97}. The X-ray emission
from the SNR is well described by a non-equilibrium
ionization (NEI) plasma model at temperature 0.3~keV with a large
(unabsorbed) flux of $\sim$10$^{-9}$ erg cm$^{-2}$ s$^{-1}$ (estimated from the
\textit{Einstein Observatory} detection; \citealt{sew90}).
The values required for other parameters in our calculations, 
including $K_{ep}$, the magnetic field strength $B$,
the average density of the target baryons (with which the energetic 
particles interact) $n_t$,
the total energy of protons $W_p$, and the total energy of 
electrons $W_e$, are summarized in Table~\ref{tab3}. 
In the $\pi^0$ decay model, 
$W_p=0.75\times 10^{50} (n_t/10\ {\rm cm}^{-3})^{-1}$ was needed.
If we constrain $W_p$ to be smaller than $50\%$ of $E_0$, where $E_0$ is
the total blast energy, we have
$n_t > 1 (E_0/10^{51}\mbox{erg})^{-1}\mbox{cm}^{-3}$.
From the X-ray spectral analysis of the SNR with the NEI 
plasma model, the density of the X-ray emitting gas $n_x$ was estimated
to be $n_x\approx 0.3\pm0.1 (E_0/10^{51}\mbox{erg})^{-1/2} \mbox{cm}^{-3}$
(\citealt{nug+84}; see also \citealt{gph97}).
In the hadronic scenario in which
the relativistic protons impact the adjacent molecular clouds, such a 
low $n_x$ value does not contradict the above large $n_t$ estimate. 
This is because the $n_t$ value includes the contribution of the baryons 
in the dense molecular clumps,
while $n_x$ reflects the low density of the interclump hot gas.
Therefore, the hadronic model seems to be consistent with the context of
shock--molecular-cloud interaction.

Based on the current \fermi\ measurements,
the cut-off energies for electrons and protons in our models are
at $\sim$1 TeV.
Thus far no very high energy (VHE) TeV detection of a source at the position
of RCW~103 has been reported (e.g., \citealt{car+13}).
If HESS J1616$-$508 is the associated TeV nebula,
in addition to the apparent positional offset, its flux at $\sim$200 GeV 
\citep{hessgal06} 
would also be slightly larger (by a factor of $\sim$2)
than the \fermi\ value we obtained.
Comparing to other CCO SNRs with ages of several thousands years, 
while the SED of Vela Jr. has a prominent TeV component
\citep{aha+07} and starts decreasing from 
above $\sim$1~TeV, resulting $>$10 TeV cut-off energies \citep{tan+11}, 
the SEDs of RCW~103 and Pup~A \citep{hew+12} are rather similar, 
as they both start decreasing above 10~GeV 
and thus are modeled to have low cut-off energies at $\sim$1~TeV
(PKS 1209$-$51/52 is not included in discussion here
since it was weakly detected by \fermi; \citealt{ara13}). 
The fact that both RCW~103 and Pup~A are in the vicinity of molecular 
clouds (e.g., \citealt{hew+12}) could be the reason for the similar 
low cut-off energies, as particles from SN explosion shocks 
is possibly limited to have relatively low energies due to the interaction with 
high-density ambient gas (e.g., \citealt{stu+97}). 
The detectability of Pup~A by the current generation 
Cherenkov telescopes has been pointed out by \citet{hew+12}, particularly 
since it is not located in a complex region and has a relatively high
Galactic latitude of $-$3\fdg4. 
We note that based on our current models, the TeV counterpart
to RCW~103 should be detectable, as the HESS survey of the Galactic plane
had a sensitivity limit of 
$\sim 10^{-12}$ erg cm$^{-2}$ s$^{-1}$ at 1~TeV \citep{hessgal06},
lower than our model fluxes at the energy.
 
\acknowledgments

We thank P. H. T. Tam for help with understanding the HESS survey
of the Galactic plane and the sensitivity.

This research was supported by 
National Natural Science Foundation of China (11073042, 11373055, 
and 11233001).
ZW is a Research Fellow of the One-Hundred-Talents project of Chinese 
Academy of Sciences.

\bibliographystyle{apj}

\begin{deluxetable}{lccc}
\tablecaption{Spatial distribution analysis results for the excess emission
at RCW 103}
\tablewidth{0pt}
\startdata
\hline
\hline
Source model & Radius & Flux\tablenotemark{a} & TS \\
\hline
Point source  & \nodata & 2.1$\pm$0.7  & 23.5 \\
Uniform disk & 0\fdg1 & 2.9$\pm$0.8 & 30.3 \\
             & 0\fdg16 & 3.5$\pm$0.9 & 36.0 \\
             & 0\fdg2 & 3.7$\pm$0.9 & 37.0 \\
             & 0\fdg3 & 4.3$\pm$1.0 & 37.5 \\
             & 0\fdg4 & 4.8$\pm$1.1 & 33.1 \\
             & 0\fdg5 & 5.1$\pm$1.2 & 28.4 \\
\enddata
\tablecomments{The analysis was made in the energy range of 30--300 GeV.}
\tablenotetext{a}{Flux is in units of 10$^{-10}$ photon cm$^{-2}$ s$^{-1}$.}
\label{tab1}
\end{deluxetable}

\begin{deluxetable}{lcc}
\tablecaption{\fermi/LAT spectral data points}
\tablewidth{0pt}
\startdata
\hline
\hline
$E$ & $E^2dN(E)/dE$ & TS \\
(GeV) & (10$^{-11}$ erg cm$^{-2}$ s$^{-1}$) & \\
\hline
1.8 & 1.67$\pm$0.34 & 55.8 \\
5.5 & 1.76$\pm$0.26 & 71.2 \\
17.3 & 1.71$\pm$0.31 & 49.7 \\
54.2 & 1.74$\pm$0.46  & 27.1 \\
169.6 & 1.14$\pm$0.63 & 6.6 \\
\enddata
\label{tab2}
\end{deluxetable}

\begin{deluxetable}{lccccc}
\tablecaption{Parameters for the hadronic and leptonic models}
\tablewidth{0pt}
\startdata
\hline
\hline
Model & $K_{ep}$ & $B$      & $n_t$ & $W_p$ & $W_e$ \\
      &          & ($\mu$G) & (cm$^{-3}$) & (erg) & (erg) \\
\hline
IC & 0.1 &  5.5    & 1   & $1.3\times 10^{50}$ & 1.1$\times 10^{49}$ \\
Brem. & 0.3 &  12   & 10   & $1.4\times 10^{49}$ & 0.35$\times 10^{49}$ \\
$\pi^0$ decay & 0.01 &  35  & 10 & $7.5\times 10^{49}$ & 0.065$\times 10^{49}$ \\
\enddata
\tablecomments{$\alpha_e=2.0$, $E_{e,cut}=1$ TeV, $\alpha_p$=2.0, 
and $E_{p,cut}=3$ TeV were used for all the models. The energy density
for the interstellar radiation field at the location was 
0.5 eV cm$^{-3}$ \citep*{pms06}.}
\label{tab3}
\end{deluxetable}

\end{document}